\documentclass[usegraphicx,useAMS,usenatbib]{mn2e}   
\usepackage{epsf}  
\usepackage{amssymb}  
\usepackage{times}  
\usepackage{aas_macros} %needed to define journal macros generated by  
 \usepackage{graphicx}  

\usepackage[usenames]{color} 
\newcommand{\paperi}{Paper~I} 
\def\dd{{\rm d}} % MC - using in equations

\newcommand{\pmf}{\pm f}

\newcommand{\vsini}{\ensuremath{\,v \sin i}}   
\newcommand{\sci}[2]{{:1}\!\times10^{:2}}      
\newcommand{\ie}{i.e.}  
\newcommand{\eg}{e.g.}

\newcommand{\teff}{$T_{\rm eff}$}  
  
\newcommand{\topp}{\emph{top}}  
\newcommand{\midd}{\emph{middle}}  
\newcommand{\bott}{\emph{bottom}}

\newcommand{\panel}{\emph{panel}}  
\newcommand{\panels}{\emph{panels}}  
  
\newcommand{\cday}{c\,d$^{-1}$}  
  
\newcommand{\mhz}{$\mu$Hz}

\newcommand{\acir}{$\alpha$\,Cir}  
\newcommand{\acirr}{$\alpha$\,Circini}  
  
\newcommand{\acirrbold}{\mbox{\boldmath $\alpha$}~Circini} % Title of paper
\newcommand{\acirbold}{\mbox{\boldmath $\alpha$}~Cir}      % Title of Section 5
\newcommand{\fonebold}{\mbox{\boldmath $f_1$}}             % Title of Section 4.2

\newcommand\kms{\ifmmode{\rm km\thinspace s^{-1}}\else km\thinspace s$^{-1}$\fi}

\newcommand{\str}{Str\"omgren}

\newcommand{\uves}{{UVES}}

\newcommand{\saao}{{SAAO}}

\newcommand{\period}{{\sc period04}}  
  
\newcommand{\hipp}{{\em Hipparcos}}  
  
\newcommand{\wire}{{\em WIRE}}  
\newcommand{\templogg}{{\sc templogg}}  
  
\def\note #1]{{\bf #1]}}

\title[Asteroseismic analysis of the roAp star \acir]  
      {Asteroseismic analysis of the roAp star \acirrbold:
84 days of high-precision photometry from 
the \wire\ satellite\thanks{Based on observations  
at the South African Astronomical Observatory.}}
    
\author[H. Bruntt et al.]  
         {H.\     Bruntt$^{1}$,  
          D.\ W.\ Kurtz$^{2}$,  
          M.\ S.\ Cunha$^{3}$,  
          I.\ M.\ Brand{\~a}o$^{3,4}$,   
          G.\     Handler$^{5}$,  
          T.\ R.\ Bedding$^{1}$, \newauthor  
          T.\     Medupe$^{6,7}$, 
          D.\ L.  Buzasi$^8$,  
          D.\     Mashigo$^{6}$,  
          I.\     Zhang$^{6}$,  
          F.\     van Wyk$^{7}$\\  
         $^1$Sydney Institute for Astronomy, School of Physics A29, University of 
Sydney, 
2006 NSW, Australia\\  
         $^2$Jeremiah Horrocks Institute of Astrophysics and Supercomputing, 
University of Central Lancashire, Preston 
PR1 2HE, UK\\  
         $^3$Centro de Astrof\'isica da Universidade do Porto, Rua das Estrelas, 
4150-Porto, Portugal\\  
         $^4$Departamento de Matem\'atica Aplicada, Faculdade de Ci\^encias, 
Universidade do Porto, 4169 Porto, Portugal\\  
         $^5$Institut f\"ur Astronomie, Universit\"at Wien, T\"urkenschanzstrasse 
17, A-1180 Wien, Austria\\  
         $^6$Department of Astronomy, University of Cape Town, Rondebosch 7700, 
South Africa\\  
         $^7$South African Astronomical Observatory, P.O.\ Box 9, Observatory 
7935, South Africa\\
         $^8$Department of Astronomy, University of Washington, Seattle WA 98195, 
USA\\ 
          }

\begin{document}  
  
\date{Accepted XXX; Received YYY; in original form ZZZ}  
  
\pagerange{\pageref{firstpage}--\pageref{lastpage}} \pubyear{2009}  
  
\maketitle  
  
\label{firstpage}

\begin{abstract}  
We present a detailed study of the pulsation of \acirr, the brightest of the  
rapidly oscillating Ap stars. We have obtained 84 days of high-precision  
photometry from four runs with the star tracker on the \wire\ satellite. 
Simultaneously, we collected ground-based Johnson $B$ observations on 16 nights  
at the South African Astronomical Observatory.
In addition to the dominant oscillation mode at $2442$\,\mhz, we 
detect two new modes  
that lie symmetrically around the principal mode to form a triplet.
The average separation between these modes is $\Delta f = 30.173 \pm 0.004$\,\mhz\
and they are nearly equidistant with the separations differing by only $3.9$~nHz.
We compare the observed frequencies with theoretical pulsation models
based on constraints from the recently determined interferometric 
radius and effective temperature, and the recently updated \hipp\ parallax. 
We show that the theoretical large separations for 
models of \acir\ with global parameters within the $1\,\sigma$ observational 
uncertainties vary between $59$ and $65$\,\mhz. 
This is consistent with the large separation being twice 
the observed value of $\Delta f$, indicating that the three main modes 
are of alternating even and odd degrees. 
The frequency differences in the triplet are significantly smaller than 
those predicted from our models, for all possible combinations of mode degrees, 
and may indicate that the effects of magnetic perturbations need 
to be taken into account. The \wire\ light curves are modulated by a 
double wave with a period of $4.479$\,d, and a peak-to-peak amplitude of 
4\,mmag. This variation is due to the rotation of the star and is a new 
discovery, made possible by the high precision of the \wire\  photometry. The 
rotational modulation confirms an earlier indirect determination of the rotation 
period.  The main pulsation mode at $2442$\,\mhz\ has two sidelobes split by 
exactly 
the rotation frequency. From the amplitude ratio of the sidelobes to 
the central peak we show that the principal mode is consistent with an oblique 
axisymmetric dipole mode ($l = 1, m = 0$), or with a magnetically 
distorted mode of higher degree with a dominant dipolar component. 
\end{abstract} 

\begin{keywords} 
stars: individual: \acir\ -- stars: oscillations -- stars: rotation -- stars: 
fundamental parameters -- stars: variables: other 
\end{keywords} 
 
% A file with the MNRAS keywords is here: ~/papers/mnras/mnras_keywords.pdf 
% ``Up to six key words are allowed for each paper'' 
 
%---------------------------------------------------------------------------% 
% 
%		 INTRODUCTION 
% 
%---------------------------------------------------------------------------% 
 
\section{Introduction} 
 
\acirr\ (HR\,5463, HD\,128898, HIP\,71908; $V=3.2$) 
is the brightest of the rapidly 
oscillating chemically peculiar A-type (roAp) stars (see reviews by 
\citealt{kurtz00} and \citealt{kurtz04}). It was among the first of the class to 
be discovered \citep{kurtz82}, showing photometric variations with a period 
$P_{\rm pul} = 6.8$\,min and an amplitude of a few mmag in Johnson $B$ 
\citep{kurtz81}. The multi-site photometric campaign by \cite{kurtz94} revealed 
four low amplitude modes in addition to the main mode. The roAp stars pulsate in 
high overtones and are described by the oblique pulsator model \citep{kurtz82, 
shibahashi93, takata95,
bigot02, saio05}. In this model the magnetic axis is inclined to the 
rotation axis. The oscillation modes follow the magnetic field, 
therefore the deformation of the star is observed from different perspectives. 
This was observed in \acir\ by \cite{kurtz94}, who found a significant 
modulation of 
the amplitude of the main mode, making it possible to determine the rotation 
period to be $P_{\rm rot} = 4.4790 \pm 0.0001$\,d. 
 
Simultaneous excitation of several high-order modes in roAp stars makes them ideal 
targets for detailed asteroseismic studies. Asymptotic theory predicts a regular 
pattern of frequencies, dominated by the large separation, which depends mainly on 
the mean density of the star. Asteroseismology of roAp stars makes it possible to 
probe their internal properties, including aspects of the structure and magnetic 
fields \citep{cunha03}. Previous detailed studies of multi-mode 
roAp stars have in some cases been limited because the fundamental parameters are 
poorly known (\eg\ \citealt{cunha07}). For example, the effective 
temperature determined from different photometric indices or spectroscopic 
analyses typically give results spanning 500\,K or more \citep{netopil08}. 
This large range may be explained, in part, 
by the chemical peculiarities and strong magnetic fields in roAp stars, 
resulting in heavy line blanketing in some regions of the spectra. 
\cite{matt99} presented observational evidence of systematic errors in the 
effective temperatures of a dozen roAp stars, based on the measurements of their 
large separations. 
 
In our companion paper \citep[\paperi]{bruntt08}, we used interferometric 
measurements with the Sydney University Stellar Interferometer (SUSI)
to determine the radius of \acir. Combined with a determination of 
the bolometric flux from calibrated spectra, we found the temperature to be 
$7420\pm170$\,K. Unlike previous determinations of \teff, which range from 7670 to 
8440\,K, our determination is nearly independent of atmospheric models -- only the 
limb-darkening correction depends on the adopted grid of model atmospheres, and 
weakly so. 
 
Here we report observations of \acir, made simultaneously with the \wire\ 
satellite and from the ground. The high-precision satellite photometry has allowed 
us to detect the large separation in \acir\ for the first time. This new result, 
combined with the constraints on the radius and \teff\ from \paperi, allow us to 
make an asteroseismic investigation based on theoretical models.

%                                                  
%--------------------------------------------------------------  
%                        OBSERVING LOG: WIRE, SAAO + UVES  
%--------------------------------------------------------------  
%                                                  
\begin{table}  
\caption{List of photometric observations from the \wire\ satellite  and the 
0.5-m and 0.75-m telescopes at \saao. 
In the last column we list the point-to-point scatter ($\sigma$) in the 
time series.}  
\label{tab:data} \centering  
\begin{tabular}{l l r r | l} \hline \hline  
 \multicolumn{1}{c}{Source}  & \multicolumn{1}{c}{Observing dates} & 
\multicolumn{1}{c}{$T_{\rm obs}$} &  \multicolumn{1}{c}{Data  } & 
\multicolumn{1}{c}{$\sigma$}   \\   
                             &                                     & 
\multicolumn{1}{c}{[days]       }  &  \multicolumn{1}{c}{points} & 
\multicolumn{1}{c}{[mmag]  }   \\ \hline  
\wire\   & 2000 Aug.\ 18--Sep.\ 29      & 42.2 &  39\,479    & 0.93  \\     
$-$      & 2005 Feb.\ 16--25            &  8.3 &  10\,857    & 1.04  \\     
$-$      & 2006 Feb.\ 10--Mar.\ 15      & 31.0 &  29\,616    & 0.91  \\     
$-$      & 2006 July 7--31              & 24.3 &  27\,273    & 0.61  \\   
\hline   
\saao\   & 2006 Feb.\ 7--Feb.\ 19       & 12.0 &  1\,802     & 2.4   \\     
$-$      & 2006 June 23--July 24        & 31.2 &  2\,330     & 2.0   \\     
\hline \end{tabular} \end{table}  
 
\begin{table}  
\caption{List of photometric observations from the 0.5-m and 0.75-m telescopes at 
\saao. The first and second columns give the initials of the observer and the 
telescope used. The duration in the fourth column is from start to end of 
observations  for each night which sometimes includes gaps due to poor weather. 
The fifth column gives the number of averaged 40-s integrations obtained and the 
last column is the standard deviation of one observation with respect to the mean 
for the night. }

\label{tab:data2}  
\centering  
\begin{tabular}{crcrrc} 
\hline 
\multicolumn{1}{c}{Obs.} & \multicolumn{1}{c}{Tel} &  
\multicolumn{1}{c}{Date} & %% \multicolumn{1}{c}{JD} &  
\multicolumn{1}{c}{$T_{\rm obs}$} & \multicolumn{1}{c}{Data} &  
\multicolumn{1}{c}{$\sigma$}  
\\ 
\multicolumn{1}{c}{} & \multicolumn{1}{c}{} &  
\multicolumn{1}{c}{2006} & %% \multicolumn{1}{c}{2400000+} &  
\multicolumn{1}{c}{[h]} &  \multicolumn{1}{c}{points} &  
\multicolumn{1}{c}{[mmag]} 
\\ 
\hline 
 
GH &  0.5-m &07--08 Feb. &  0.58  &   49  &  3.5  \\ 
GH &  0.5-m &12--13 Feb. &  0.95  &   83  &  3.4  \\ 
GH & 0.75-m &13--14 Feb. &  3.04  &  258  &  1.8  \\ 
GH &  0.5-m &13--14 Feb. &  0.45  &   40  &  2.0  \\ 
GH & 0.75-m &14--15 Feb. &  3.97  &  341  &  2.4  \\ 
GH & 0.75-m &15--16 Feb. &  2.58  &  224  &  1.8  \\ 
GH & 0.75-m &16--17 Feb. &  3.95  &  264  &  2.3  \\ 
GH & 0.75-m &17--18 Feb. &  3.49  &  302  &  2.7  \\ 
GH & 0.75-m &19--20 Feb. &  2.88  &  241  &  2.5  \\ 
total  &    &            & 21.89  & 1802  &        \\ 
\\                                                 
DM &  0.5-m &23--24 Jun. &   5.45 &  381  &  2.0  \\ 
DM &  0.5-m &24--25 Jun. &   5.45 &  445  &  1.7  \\ 
DM &  0.5-m &07--08 Jul. &   5.22 &  418  &  1.9  \\ 
DM &  0.5-m &11--12 Jul. &   3.89 &   88  &  2.1  \\ 
DM &  0.5-m &17--18 Jul. &   4.79 &  396  &  1.8  \\ 
DM &  0.5-m &18--19 Jul. &   4.79 &  389  &  2.0  \\ 
IZ &  0.5-m &24--25 Jul. &   4.16 &  213  &  1.9  \\ 
total &     &            &  33.75 & 2330  &        \\

\hline  
\end{tabular}   
\end{table}

%                                                  
%--------------------------------------------------------------  
% Program: .r wire_acir_plot_lc.pro  
   \begin{figure}  
   \centering  
   \includegraphics[width=8.7cm]{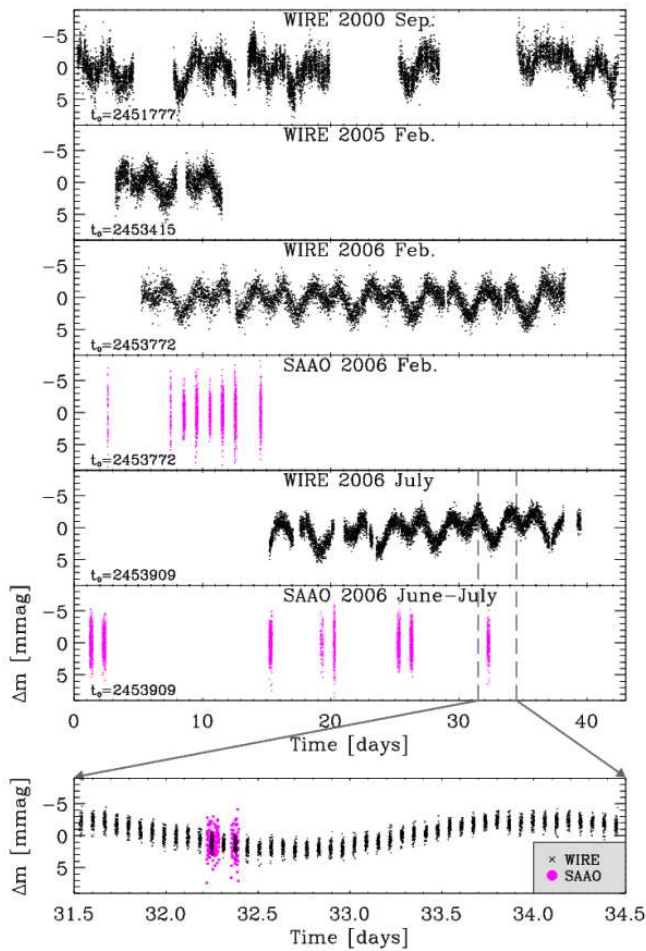}  
 \caption{The four light curves of \acir\ from \wire\ and the two light curves 
from \saao. 
The different zero points in time are indicated in each panel. 
The \bott\ \panel\ shows an expanded view of three days where simultaneous 
observations with \wire\ and \saao\ were made. } 
         \label{fig:lc}  
   \end{figure}  
%  
%______________________________________________________________  

%---------------------------------------------------------------------------%  
%  
%		      OBSERVATIONS AND DATA REDUCTION  
%  
%---------------------------------------------------------------------------%  

\section{Observations}\label{sec:obs} 
 
Soon after the launch of the \wire\ satellite on 4 March 1999,
the hydrogen coolant for the main infrared camera was lost. However, the star 
tracker was used to observe around 240 stars with apparent magnitudes $V<6$ until 
communication with the satellite failed on 2006 October 23 \citep{buzasi02, 
bruntt07, 
bruntt07gott}. 
The 52\,mm aperture illuminated a SITe CCD camera and
windows centered on the five brightest stars in the field were acquired
at a cadence of 2~Hz. Each \wire\ field was typically monitored for 2 to 4 weeks 
and 
some targets, including \acir, were observed in more than one season. \wire\ 
switched between two targets during each orbit in order to best avoid the 
illuminated face of the Earth. Consequently, the duty cycle was typically around 
$20$ to $40$ percent, or about $19$ to $38$ minutes of continuous observations per 
orbit. The time baseline of the observations analysed here is almost six years, 
during which the orbital period of the satellite decreased from $96.0$ to 
$93.3$ minutes. 
 
The roAp star \acir\ was observed during four runs with \wire. 
The first data set in 2000 September was not optimal in that several long gaps are 
present and the sub-pixel position on the CCD was less stable than the succeeding 
runs. The second run in 2005 February lasted 8 days, and was followed by runs 
of 31 and 24 days in 2006 February and July. A log of the observations is given in 
Table~\ref{tab:data}. Excluding the gaps in the time series, \acir\ was observed 
for 84 days, making it one of the most observed stars with \wire. 
 
To be able to compare the \wire\ results with previous ground-based studies,
we collected additional ground-based photometric time-series. 
During both runs with \wire\ in 2006 we made simultaneous 
photometric observations in the Johnson $B$ filter from the South African 
Astronomical Observatory (\saao). 
We used the Modular Photometer on the 0.5-m 
telescope and the University of Cape Town Photometer on the 0.75-m telescope. All 
observations were made through a Johnson $B$ filter, plus a neutral density filter 
of 2.3\,mag on the 0.5-m telescope and 5\,mag on the 0.75-m telescope. The neutral 
density filters were needed to keep the count rates for this very bright star at 
levels that would not damage the photomultiplier tubes, or result in large 
dead-time corrections. The resultant count rates were about 
380\,000\,s$^{-1}$ for the 0.5-m telescope and 
150\,000\,s$^{-1}$ for the 0.75-m telescope. 
A detailed log of the observations from \saao\ is given in Table~\ref{tab:data2}. 
 
The light curves from \wire\ and \saao\ are shown in Fig.~\ref{fig:lc}. 
The \bott\ \panel\ shows an expanded view of simultaneous observations. 
In each panel we give the time zero point $t_0$ 
as the Heliocentric Julian Date (HJD). 

\subsection{Data reduction\label{sec:reduction}} 
 
The \wire\ data set consists of $3.3$ million windows extracted by the on-board 
computer from the 512$\times$512 pixels CCD star tracker camera. Each window is 
8$\times$8 pixels and aperture photometry was carried out using the pipeline 
described by \cite{bruntt05}. 
The data collected within 15 seconds are binned to decrease 
the number of data points to $107\,225$ (cf.\ Table~\ref{tab:data}).
We note that \acir\ is a visual binary with a separation of 
15\rlap.$^{\prime\prime}$6. Since the pixel size of the \wire\ CCD is 
60$^{\prime\prime}$, the B~component is entirely within the applied photometric 
aperture. As noted by \cite{kurtz94} the contamination is negligible, since the 
difference in magnitude is $\Delta V = 5.05$. With the \str\ indices of the 
B-component from \cite{sina89} and using the \templogg\ software 
\citep{rogers95,kupka01} for the calibration, we find \teff\,$\simeq 
4660\pm200$\,K, 
$\log g = 4.6\pm0.2$, and ${\rm [Fe/H]}=-0.26\pm0.10$, suggesting that it is a 
K5\,V star. In the following, we use \acir\ to mean the A~component of the system. 

The raw light curves contain a small fraction ($1$ 
to $10$\%) of outlying data points. These points were identified and removed based 
on the measured position, background level, and width of the stellar image (PSF). 
The light curves from Sep'00, Feb'05 and Feb'06 have a high amount of scattered 
light towards the end of each orbit, increasing the background level from about 20 
to 1400 ADU. In these light curves we found a significant correlation between the 
background and the measured flux, which we removed by a spline fit. The light 
curve from Jul'06 has low background, and in this case we found a correlation of 
the measured flux and the width of the PSF, which was removed by subtracting a 
second-order polynomial. 
 
The \saao\ data were obtained in high-speed mode using continuous 10-s 
integrations. With continuous observations of the target star, no comparison stars 
were observed. Sky observations were made as needed, based on conditions 
throughout the observing run. The data were corrected for dead-time losses, sky 
subtracted and the mean extinction was removed. They were then averaged four 
points at 
a time into 40-s integrations. Some remaining low frequency variability, caused by 
transparency variations, was removed with high-pass filtering that did not affect 
the pulsation frequencies of interest in the data. This latter step results in 
white noise across the frequency spectrum, and hence better estimates of errors on 
amplitudes and phases than would be the case if low frequency variance were left 
in the data. The last column of Table~\ref{tab:data2} gives the standard 
deviation for each night of observations. 
The contribution to the standard deviation from the actual pulsations is small, 
so it a good measure of the quality of the night. 
On some nights, weather interruptions resulted in gaps in the data, 
which is reflected in the number of data points 
and duration of each night in Table~\ref{tab:data2}.

%---------------------------------------------------------------------------% 
% 
%	 LIGHT CURVE ANALYSIS		 
% 
%---------------------------------------------------------------------------% 
 
\section{Light curve analysis\label{sec:lc}} 
 
The light curves of \acir\ show variations on two different time scales. 
The double wave seen in the \wire\ data in Fig.~\ref{fig:lc} has a period of 
$4.479$\,d 
and is due to rotation, seen here for the first time,
while the periods of the rapid oscillations lie in the 
range $6.5$ to $7.0$\,min, or three orders of magnitude faster. In the following 
Sections we will analyse these variations in detail. 
 
\subsection{Rotational period and inclination angle} 
\label{sec:rot} 

\begin{figure} 
 \centering 
 \includegraphics[width=8.5cm]{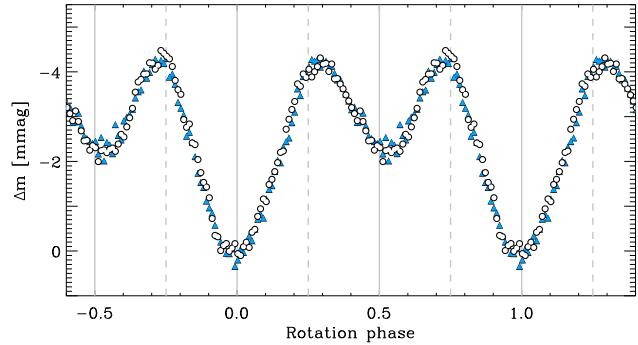} 
 \caption{Binned light curves of \acir\ from \wire\ phased with the rotation 
period. Triangle symbols are data from Feb'06 and circles are from Jul'06. 
Vertical lines are shown to guide the eye. The zero point in time for the phase is 
$t_{{\rm rot}} = $\,HJD245\,3937.2086. 
\label{fig:rotation}} 
 \end{figure}

To measure the rotation period, we fitted the observed double wave
with two sinusoids with frequencies $f_A$ and $f_B = 2 \times f_A$, 
yielding a period of $P_{\rm rot} = 4.4792\pm0.0004$\,d. 
In this analysis we first subtracted the high-frequency 
oscillations (see Sect.~\ref{sec:hifreq}). In Fig.~\ref{fig:rotation} we show the 
binned lights curves from Feb'06 and Jul'06, phased with the rotation period with 
a zero point in time $t_{{\rm rot}} = $\,HJD245\,3937.2086. The peak-to-peak 
variation is only about $4$\,mmag and it is the excellent precision and 
low frequency stability of the \wire\ data that have allowed us to make the first 
direct measurement of the rotation of \acir. Ground-based observations have not 
been able to do this, partly because of the low amplitude of the variations and 
partly because of the challenge of making differential photometric observations on 
such a bright star. 
 
From the rotation period and the interferometrically determined radius of 
$R/{\rm R}_\odot = 1.967\pm0.066$ (\paperi), we find the rotational velocity to be 
$v_{\rm rot}=21.4\pm0.7$\,\kms. We can thus estimate the inclination of the 
rotation axis from the measured \vsini. The average value from \cite{kupka96} and 
\cite{balona03} is $v \sin i = 13.0\pm1.5$\,\kms. 
A more accurate value has been determined from spectra collected 
with the \uves\ spectrograph: $v \sin i = 13.0\pm1.0$\,\kms\ (Vladimir Elkin, 
private communication). 
These \uves\ spectra have better resolution and higher S/N than previous 
studies, thus we use this value of $v \sin i$ 
to determine the inclination to be $i = 37 \pm 4^\circ$.

The rotation of \acir\ was first measured indirectly by \cite{kurtz94} from the 
amplitude modulation of the principal oscillation mode at 2442\,\mhz. 
From the observed frequency separation of the two symmetrical side-lobes, 
\cite{kurtz94} determined the period $P_{\rm rot}=4.463\pm0.010$\,d. 
This was refined to get $P_{\rm rot}=4.4790\pm0.0001$\, 
using photometry from several observing campaigns with a baseline of 12\,y 
\citep{kurtz94}.
These results are in very good agreement with our direct measurement. 
 
% 
 %-------------------------------------------------------------- 
 % Program: .r wire_acir_phaseplot (use the second plot made in that program) 
 
 % 
 %______________________________________________________________ 

We interpret the modulation of the light curves in Fig.~\ref{fig:rotation}
as being due to surface features, 
most likely two spots of over-abundances of rare earth elements located near the 
equator, thus not associated with the magnetic and pulsation poles. Such spots 
are typical of the $\alpha^2$\,CVn class of photometrically 
variable magnetic Ap stars (see, e.g., \citealt{pyper69}). Redistribution of flux 
from blue to red because of the increased line blanketing in the blue often gives 
rise to rotational light variations that are out-of-phase between blue- and 
red-filter observations in Ap stars. We cannot determine with certainty whether 
this 
is the case for \acir, as we have only very broad-band observations with \wire\ 
with an uncertain filter function (see section~\ref{sec:saao}). Since the flux 
from the star is dominated by blue light, given that $T_{\rm eff} = 7420\pm170$\,K 
(\paperi), it is likely that the double wave variation shows minima when 
each spot is closest to the line-of-sight. The two spots are of different size, 
hence giving the double wave variation with rotation. 
We note that the double wave is 
asymmetrical (as can be seen in Fig.~\ref{fig:rotation}), since the 
secondary minimum lies at phase $0.519\pm0.008$. It is worth noting that the 
configuration of the spots that give rise to the rotational light modulation did 
not appear to change significantly during any of the runs with \wire. We will 
discuss the spot configuration in more detail in a future publication concerning 
Doppler imaging of the rotational light curve. 
  
Measurements of the magnetic field of \acir\ are inconclusive regarding its 
variability with the rotation period. \cite{bychkov05} reanalysed sparse data 
collected from the literature spanning 14\,y and showed a rotation curve 
consistent with $P_{\rm rot} = 4.4794$\,d. We have repeated that exercise 
including a new measurement (\citealt{hubrig04}), making the time span 23\,y, and 
find a marginal indication of magnetic variability with the known rotation period 
$P_{\rm rot} = 4.4794$\,d, at the 2.5$\sigma$ significance level, if we neglect 
one outlying positive field measurement. This does not give us confidence that 
magnetic variability with the rotation period has actually been detected, and it 
is certainly not possible to give a time of magnetic maximum that can be compared 
to the time of rotational light minimum, or to the time of pulsation maximum. 
 
We can conclude that the longitudinal field for \acir\ has been detected. The mean 
value derived from the data of \cite{hubrig04} is the most reliable at $- 239 \pm 
27$\,G (internal error). The evidence from the data in the literature is that the 
longitudinal field is probably not reversing; i.e.\ only one magnetic pole is seen 
over the rotation period, implying that $i + \beta < 90^{\circ}$, where $i$ is the 
rotational inclination and $\beta$ is the magnetic obliquity of the negative pole. 
For oblique magnetic 
rotators, the strength of the measured longitudinal magnetic field is proportional 
to $\cos \alpha$, where $\alpha$ in the angle between the magnetic pole and the 
line-of-sight. That varies as 
\begin{equation} 
\cos \alpha = \cos i \cos \beta + \sin i \sin \beta \cos \Omega t, 
\end{equation} 
\noindent where $\Omega$ is the rotational frequency.

Our rotational phase zero -- the time of rotational light minimum in 
Fig.~\ref{fig:rotation} -- is therefore the time when the largest of the two 
equatorial spots is on the observed 
hemisphere closest to the line-of-sight. Since two equatorial spots are needed to 
describe the light curve, they cannot be associated with the one magnetic pole 
that is always visible. The pulsation pole is also not associated with the spots, 
as we will see in Sect.~\ref{sec:sidelobes}. 
This is not normal for an roAp star, hence for all 
interpretation related to the rotational light curve, we proceed 
with caution.

% ---------------------------------------------------------------------- 
\begin{table} 
\caption{Frequencies extracted from the two combined \wire\ data sets from 2006. 
Three 
frequencies were fitted to the \saao\ $B$-filter photometry as described in the 
text. The phases are with respect to $t_0 =$\,HJD245\,3939.
\label{tab:freq}} 
\centering  
\setlength{\tabcolsep}{4pt} % narrow table: default is tabcolsep = 6pt  
\begin{tabular}{l|lll|ll} \hline \hline  
   & \multicolumn{3}{c|}{\wire} & \multicolumn{2}{c}{\saao} \\ \hline  
   &  $f$    & $A_{\rm WIRE}$ & $\phi_{\rm WIRE}$ & $A_B$ & $\phi_B$ \\   
ID &  [\mhz] & [mmag]         & [0..1]   & [mmag]      & [0..1]   \\ \hline   
  
$f_1$ & $ 2442.0566(2)$   & $ 0.652(5)$ & $ 0.506(2)$  &   $1.54(3)$  &   
$0.490(3)$\\  
$f_2  $&$ 2265.43(2)$     &     $-$     &    $-$       &      $-$     &  $-$        
\\  
$f_3  $&$ 2341.79(1)$     &     $-$     &    $-$       &      $-$     &  $-$        
\\  
$f_4$ & $ 2366.47(4)$     & $ 0.047(8)$ & $ 0.67(4)$   &      $-$     &  $-$        
\\  
$f_5$ & $ 2566.98(3)$     & $ 0.044(7)$ & $ 0.18(3)$   &      $-$     &  $-$        
\\ %%% f_5 is from Jul'06 only!  
$f_6$ & $ 2411.8822(9)$   & $ 0.149(5)$ & $ 0.471(6)$  &   $0.28(3)$  &  $0.46(2)$  
\\  
$f_7$ & $ 2472.2272(5)$   & $ 0.188(5)$ & $ 0.041(4)$  &   $0.42(3)$  &  $0.06(2)$  
\\ \hline  
$f_1^-$&$ 2439.47(2)$     & $ 0.089(5)$ & $ 0.06(1)$   &      $-$     &  $-$        
\\  
$f_1^+$&$ 2444.64(2)$     & $ 0.079(5)$ & $ 0.97(2)$   &      $-$     &  $-$        
\\ \hline  
  
\multicolumn{6}{l}{Notes: $f_2$ and $f_3$ were detected by \cite{kurtz94}, but are  
not} \\  
\multicolumn{6}{l}{present in the \wire\ data. $f_5$ was only detected in the  
Jul'06 data.}\\  
\multicolumn{6}{l}{$f_1^+$ and $f_1^-$ correspond to $f_1\pmf_{\rm rot}$, where  
$f_{\rm rot}$ is the rotational}\\  
\multicolumn{6}{l}{frequency.}  
  
\end{tabular} \end{table}  
  
%  
% Program: .r wire_acir_clean_progress  
   \begin{figure*}  
\centering  
 \includegraphics[width=18.0cm]{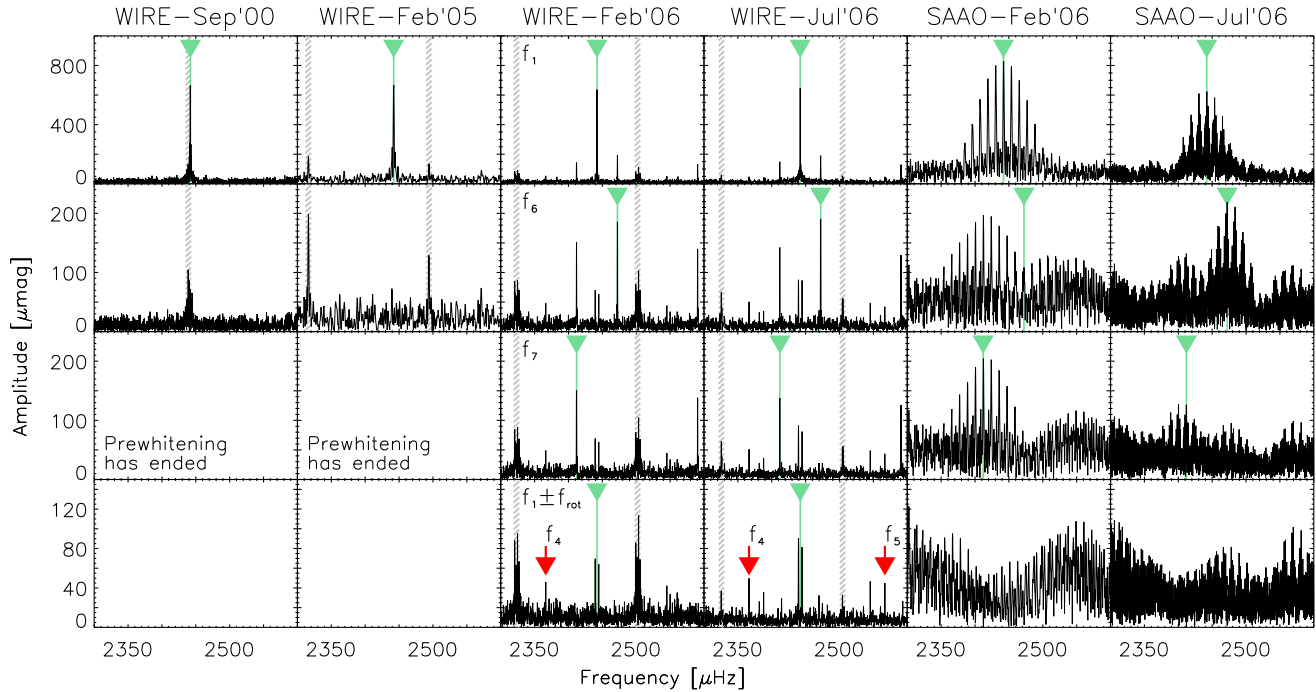} 
 \caption{Prewhitening of the \wire\ and \saao\ data sets. In each row the 
triangles and vertical lines mark the frequency being subtracted and the hatched 
grey regions 
mark the harmonics of the \wire\ orbit. In the amplitude spectra from 2000 and 
2005 we detect only the dominant $f_1$ mode. All four amplitude spectra from 2006 
show $f_1$, $f_6$ and $f_7$, although for the \saao\ data the S/N is poor for the 
two weaker modes. In the bottom row four additional peaks, $f_1\pmf_{\rm rot}$, 
$f_4$, and $f_5$, are marked in the \wire\ data sets from 2006. 
\label{fig:clean1}} 
\end{figure*} 
 
\subsection{Analysis of the rapid oscillations} % modes near 2.4\,\milhz} 
\label{sec:hifreq} 
 
We analysed the rapid oscillations in the \wire\ and \saao\ light curves 
using the Fourier analysis package \period\ \citep{period}
and a Discrete Fourier Transform \citep{kurtz85}, 
and we found excellent agreement between the two methods.
We first subtracted the rotation signal, as 
found above. We initially analysed each data set independently, and the 
prewhitening process is shown in Fig.~\ref{fig:clean1}. The \topp\ \panels\ show 
the amplitude spectra of the six observed light curves, \ie\ four from \wire\ and 
two from \saao. In each row of panels the frequency with the highest amplitude is 
marked by the triangle and a vertical line. The hatched regions mark the harmonics 
of the orbital 
frequency of \wire, 
which changed considerably from Sep'00 ($f_{\rm WIRE}=174.2$\,\mhz) to 
Jul'06 ($f_{\rm WIRE}=178.9$\,\mhz) due to the decay of the orbit. The signal seen 
at the orbital harmonics ($n \times f_{\rm WIRE}$) has leaked from drift noise at 
low frequencies. Note that in the Sep'00 spectrum, the main $f_1$ mode lies close 
to an orbital harmonic ($n=14$). The second row of panels in Fig.~\ref{fig:clean1} 
shows the residual amplitude spectra after subtracting the $f_1$ mode at 
$2442$\,\mhz. No additional significant peaks are detected in the \wire\ data from 
Sep'00 and Feb'05. However, both the \wire\ data sets from 2006 and the 
simultaneous \saao\ data show two peaks, $f_6$ and $f_7$, that have not been 
detected in previous campaigns. Interestingly, these peaks lie symmetrically at 
$\pm30$\,\mhz\ around the main mode. Note that in the amplitude spectra from 
\saao, the 1\,\cday\ alias peaks are sometimes slightly higher than 
the ``true peak'' (known from the \wire\ data and marked by the vertical line). 
This problem is well known for single-site data, especially at low S/N. 
Fortunately, there is no ambiguity in our case since we have the high-S/N 
satellite data. 
In the \bott\ \panels\ in Fig.~\ref{fig:clean1} we see no signal above the noise 
in either of 
the \saao\ data sets. In the Feb'06 and Jul'06 \wire\ data sets we detect 
additional peaks\footnote{$f_{\rm rot}=1/P_{\rm rot}=2.5841\pm0.0001$\,\mhz.} at
$f_1^{\pm}=f_1\pmf_{\rm rot}$ and $f_4=2366$\,\mhz, while in Jul'06 we detect also 
a low amplitude peak at $f_5=2566$\,\mhz. 
 
%-------------------------------------------------------------- 
% Program: .r wire_acir_window 
 \begin{figure*} 
 \centering 
 \includegraphics[width=13cm]{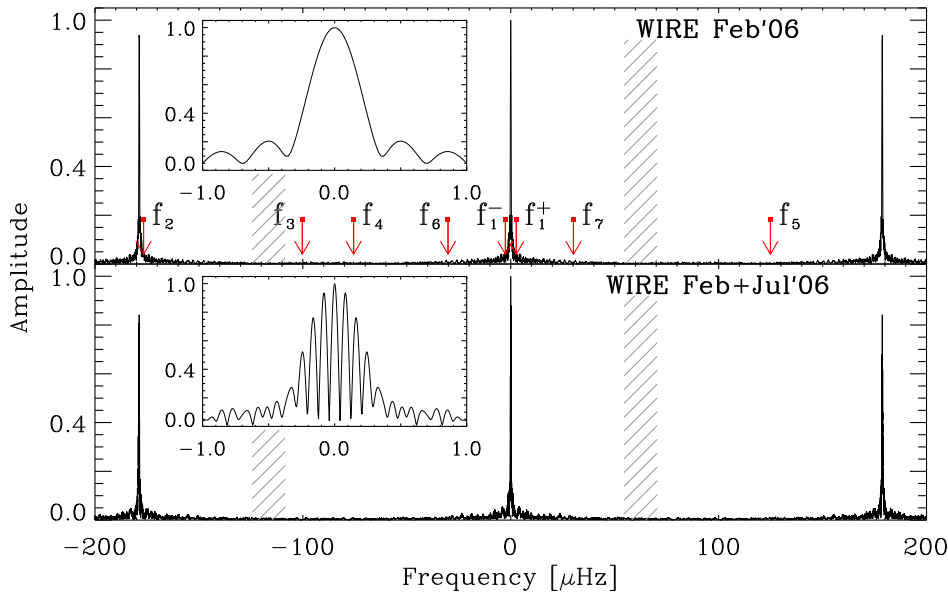} 
 \caption{Spectral windows for the Feb'06 data set (\topp\ \panel) and the 
combined 
Feb'06 $+$ Jul'06 data sets (\bott\ \panel) from \wire. The insets show details of 
the central peak. The arrows mark the positions of the frequencies detected in 
\acir\ and the hatched regions mark the location of the orbital harmonics. Note 
that $f_2$ (close to an alias peak of $f_1$) and $f_3$ are from Kurtz et al.\ 
(1994) and were not detected in the \wire\ data sets. \label{fig:win}} 
 \end{figure*} 
 
\subsection{Combining the 2006 \wire\ data sets} 
\label{sec:combine}
 
Since the two \wire\ data sets from 2006 contain nearly the same peaks
in the amplitude spectrum, 
we combined the data sets to be able to extract more precise frequencies: the 
formal 
error on frequency scales with $1/T_{\rm obs}$, where $T_{\rm obs}$ is the 
baseline of the observations (\eg\ \citealt{mont99}). There is a caveat when 
combining these data due to the 4-month gap between them, resulting in a spectral 
window with many sub-peaks that have almost equal amplitude. 
To illustrate the problem, spectral windows are 
shown in Fig.~\ref{fig:win} for the Feb'06 data set and the combined Feb'06 $+$ 
Jul'06 data sets. In the \topp\ \panel\ the detected frequencies are marked by 
arrows, including $f_2$ and $f_3$, which were observed by \cite{kurtz94} but are 
not present in the \wire\ data sets. 
 
To see if there is any ambiguity when extracting the frequencies, we repeated the 
analysis for 100 simulations of the combined 
Feb'06 $+$ Jul'06 time series, to see how often the input 
frequencies were correctly identified. To make the simulations, we followed the 
approach of \citet[their Appendix~B]{bruntt07m67}. The simulations include the 
observed frequencies, the rotational double wave, white noise, and a $1/f$ noise 
component consistent with the observations. The simulations were analysed by an 
automatic prewhitening procedure. The inserted frequencies were retrieved in all 
100 cases, except 11/100 for $f_4$ and 3/100 for $f_1^+$ (these modes have ${\rm 
S/N} \sim 6$ and $10$). In these cases frequencies offset by $\pm0.083$\,\mhz\ 
were picked by the automatic procedure, corresponding to an alias peak offset by 
$\pm1/T_{\rm gap}$, where $T_{\rm gap}$ is the duration of the gap. Due to these 
relatively high false-alarm ratios, we will quote results for the low amplitude 
modes using only the best data set from Jul'06. 
 
In Table~\ref{tab:freq} we give the final results for the combined data set for 
$f_1$, $f_6$ and $f_7$, while the results for the other modes are based on the 
Jul'06 data set. Note that the frequencies $f_1^{\pm}$ were fixed, 
with $f_1$ from the combined data set 
and $f_{\rm rot} = 1/P_{\rm rot}$ from Sect.~\ref{sec:rot}. 
We used the results of the simulations to determine the 
uncertainties of the frequency, amplitude, and phase calculated as the standard 
deviation of the extracted parameters. The uncertainties are given in the 
parentheses in Table~\ref{tab:freq}, and are in general agreement with theoretical 
expressions, assuming pure white noise (\eg\ \citealt{mont99}).

\subsection{Calibration of the \wire\ photometry\label{sec:saao}} 

To calibrate the measured amplitude and phases of the \wire\ photometry, 
we analysed the combined ground-based Johnson $B$ data from the two \saao\ runs, 
obtained at the same time as the two \wire\ data sets from 2006. 
We made simulations of the data as described in Sect~\ref{sec:combine}, 
which showed that the automatic prewhitening was seriously 
affected by the complicated spectral window with strong 1\,c\,d$^{-1}$  
alias peaks. 
Fortunately, the frequencies of $f_1$, $f_6$ and $f_7$ are determined very 
accurately from the \wire\ data and we therefore held the frequencies fixed 
when determining their amplitudes and phases. 
The fitted values are listed in the last two columns in Table~\ref{tab:freq}. 
 
We measured the amplitude ratio and the phase shift between the \wire\ and \saao\ 
observations. The amplitude ratios for the three modes $f_1$, $f_6$ and $f_7$ are: 
$A_{B} / A_{\rm WIRE} = 2.36\pm0.11$, $1.88\pm0.38$, and $2.23\pm0.36$, with 
a weighted mean $\langle A_{B} / A_{\rm WIRE} \rangle = 2.32\pm0.10$. 
The results for the $f_1$ mode shows a small but significant phase shift of 
$\phi_{\rm 
WIRE} - \phi_B = +5\rlap. ^\circ8 \pm 1\rlap.^\circ3$. The phases of $f_6$ and 
$f_7$ are in agreement with this shift, but the uncertainties are three times 
larger and cannot be used to constrain the phase shift. 

\cite{kurtz84} made simultaneous observations in Johnson $B$ and $V$ and measured 
only small phase 
shifts $\phi_V-\phi_{B-V}=+13^\circ \pm 6^\circ$ and $\phi_V - \phi_B = - 
7\rlap.^\circ4 \pm 5\rlap.^\circ 1$. They measured the amplitude ratio for 
$f_1$ to be $A_B / A_V = 2.28\pm0.26$, 
which is essentially the same as our ratio of $A_{B} / A_{\rm WIRE}$.
Thus the amplitude in the \wire\ data is similar to that which would 
be observed through a Johnson $V$ filter, 
strongly suggesting that the bandpass of the \wire\ photometry 
is centered at about the same wavelength as Johnson $V$. 
The interpretation of the small phase shifts is more problematic, 
since it is well-known from high resolution spectroscopic studies that the 
pulsation phase in roAp stars can be a sensitive function of atmospheric depth of 
the observations (see, e.g., \citealt{kurtz06b},  \citealt{ryabchikova07}).  

\section{Identification of the primary mode}
\label{sec:sidelobes} 

In the following we will discuss the possible mode identification 
of the primary $f_1$ mode using the relative amplitudes of the rotational
sidelobes. This is done in the framework of the oblique pulsator model,
which postulates that the axis of the non-radial pulsation 
modes is inclined to the rotation axis of the star. In some versions of the model 
the pulsation axis coincides with the magnetic axis (\citealt{kurtz82}, 
\citealt{shibahashi86}, \citealt{kurtz86}, \citealt{shibahashi93}, 
\citealt{takata95}), while in another version the pulsational axis may be inclined 
to both the rotational and magnetic axes \citep{bigot02}. Here, we interpret the 
rotational frequency triplet $f_1\pmf_{\rm rot}$ in terms of the formalism of 
\cite{shibahashi86} and \cite{kurtz86}.

We expect the observed modes in photometry 
to be of low degree (typically $l \le 2$), given that 
generally perturbations associated with modes of higher degree cancel out when 
averaged over the stellar disk. However, this may not always be the case in roAp 
stars, 
since the presence of strong magnetic fields that permeate the surface of these 
stars may result in a distortion of the eigenfunctions at the surface 
\citep{dziembowski96,cunha00,saio04}. Thus, a mode of degree higher than 
$l =2$ could 
in principle give rise to lower-degree components at the surface and still be 
detected. If the magnetic perturbations are symmetric about the magnetic equator, 
as is the case for pure dipolar or pure quadrupolar magnetic topologies, the 
additional components of the perturbed eigenfunctions will have the same 
parity (odd or even $l$) as the non-perturbed one. 
We will discuss this further in the third paper of this series 
(Brand\~ao et~al., in preparation), in which detailed magnetic 
modelling of the pulsations in \acir\ will also be presented.

\begin{figure} 
\centering 
\includegraphics[width=8.8cm]{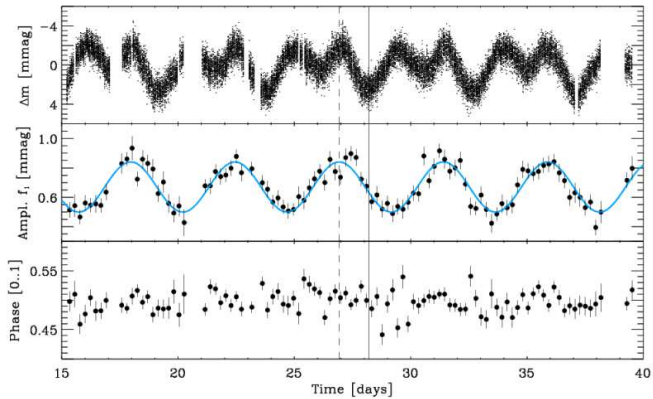} 
\caption{The \topp\ \panel\ is the light curve of \acir\ from \wire\ in Jul'06. 
The \midd\ and \bott\ \panel\ shows variation of the amplitude and phase of the 
primary mode at 2442\,\mhz\ determined by fits to subsets of the light curve. 
The vertical lines mark the time zero points 
at maximum amplitude ($t_{{\rm max}}$; dashed line) and 
minimum light ($t_{{\rm rot}}$; solid line).
\label{fig:ampvar}} 
\end{figure}

\subsection{Properties of the rotational sidelobes
\label{sec:prop}}

We observed two frequency sidelobes around the principal 
pulsation frequency~$f_1$, separated by 2.58\,$\mu$Hz,
which is the same as the rotational frequency. %%% (cf.\ Table~\ref{tab:freq}).
In the two amplitude spectra from \wire~2006 in Fig.~\ref{fig:clean1}
the sidelobes at $f_1 \pm f_{\rm rot}$ are clearly seen.
To measure the amplitudes and phases of the sidelobes
we assume they are separated by exactly the rotation frequency, 
as required by the oblique pulsator model. 
We use the highest precision value we have for the rotational frequency 
(cf.\ Sect.~\ref{sec:rot}; $f_{\rm rot} = 2.5841 \pm 0.0001$\,$\mu$Hz) 
and fitted an equally spaced frequency triplet about $f_1 = 2442.0566$\,$\mu$Hz 
by least squares,
with the requirement that the zero point of the time scale be chosen such that the 
phases of the rotational sidelobes be equal (i.e. $\phi_1^{+} = \phi_1^{-}$), as 
expected for axisymmetric modes in the oblique pulsator model \citep{kurtz82}.

The results of that fit are shown in Table~\ref{tab:sidelobes} for 
the zero point in time $t_{{\rm max}} =\,{\rm HJD}245\,3935.91562$.
The amplitudes of the two rotational sidelobes are equal to within $1.5\,\sigma$, 
i.e.\ ${A_{1}^{+}} / {A_{1}^{-}} = 0.854 \pm 0.096$.
This is consistent with the magnetic perturbations to the eigenfrequencies being 
very 
much stronger than the rotational perturbations. 
The zero point in time for the least-squares fit was 
chosen such that the phases of the rotational sidelobes are equal, but the phase 
of the central frequency of the triplet was not constrained by this choice of 
$t_{{\rm max}}$. 
The nearly identical phases (they differ by $1.6\,\sigma$) 
are thus consistent with a purely geometrical amplitude modulation 
with rotational aspect in the oblique pulsator model. 

% ----------------------------------------------------------------------------   
\begin{table} 
\caption{A least squares fit to all of the \wire\ 2006 data of the principal 
frequency, $f_1 = 2442.0566$\,$\mu$Hz and its two rotational sidelobes 
which -- within the oblique pulsator model -- are separated 
by exactly the rotation 
frequency of $f_{\rm rot} = 2.5841 \pm 0.0001$\,$\mu$Hz. \label{tab:sidelobes}} 
\centering  
\begin{tabular}{llll@{}l} 
 
\hline \hline  
   &  $f$      & $A$          & \multicolumn{2}{l}{$\phi$}    \\   
ID &  $\mu$Hz  & mmag         & \multicolumn{2}{l}{$[0..1]$}  \\  
 
\hline   
  
$f_1^-$&$ 2439.4725$ & $ 0.082 \pm 0.006$ & $ 0.423  $&$ \pm 0.011$         \\  
$f_1$ & $ 2442.0566$ & $ 0.618 \pm 0.006$ & $ 0.4050 $&$ \pm 0.0014$        \\  
$f_1^+$&$ 2444.6407$ & $ 0.070 \pm 0.006$ & $ 0.423  $&$ \pm 0.013$         \\

\hline  
\end{tabular}  
\end{table}  
% ----------------------------------------------------------------------------  
  
To investigate the variation of the amplitude
we followed the approach of \cite{kurtz94} to analyse the \wire\ 
Jul'06 data set, which has the highest S/N of all of our data. 
We first subtracted the other modes from the data set, 
\ie\ the rotational double wave $f_A + f_B$ (Sect.~\ref{sec:rot})
and the low amplitude modes $f_4$--$f_7$ (Sect.~\ref{sec:hifreq}).
We then split the data into subsets containing 30 
pulsation cycles of $f_1$, corresponding to time bins of $\Delta t \simeq 3.4$\,h. 
In each subset we found the amplitude and phase of the $f_1$ mode by least 
squares, holding the frequency fixed. 
We find that the amplitude of $f_1$ is modulated
by the rotation while the phase is not. The result is shown in 
Fig.~\ref{fig:ampvar},
where the variations in amplitude and phase of $f_1$ are shown 
in the middle and bottom panels, respectively.
The top panel shows the \wire\ light curve for comparison.
The time of minimum light ($t_{{\rm rot}}$, as defined in Sect.~\ref{sec:rot})
is marked with a solid line 
and the time of maximum amplitude ($t_{{\rm max}}$) 
is marked with a dashed line.
The delay in the maximum amplitude with 
respect to the zero point for the rotation phase (Sect.~\ref{sec:rot}) is 
$\Delta \phi = \phi_{\rm max} - \phi_{\rm rot} = 0.287\pm0.003$. 
In other words, the time of maximum amplitude occurs at rotation phase 
$1 - \Delta \phi = 0.713$ (see Fig.~\ref{fig:rotation}). 
Comparison of the 
time zero points
for the rotational light variation 
and the pulsation maximum gives nearly the same result: 
the rotational phase is
$\phi_{\rm rot} = (t_{{\rm max}} - t_{{\rm rot}})/P_{\rm rot} = -0.289 = 0.711$.

%\footnote{The zero points $t_{{\rm rot}}$ and $t_{{\rm max}}$ are 
%defined in Sects.~\ref{sec:rot} and \ref{sec:prop}.}
%  t_{\rm max} = {\rm HJD}245\,3935.9254 \simeq t_{{\rm max}}$

For an axisymmetric dipole mode ($l=1, m=0$) 
aligned with the magnetic field, and for spots 
centred on the magnetic poles, we expect the time of pulsation maximum to coincide 
with the time of rotational light minimum, whereas, as can be seen in 
Fig.~\ref{fig:ampvar}, rotational phase 0.71 coincides with one of the rotational 
light maxima. The essentially pure amplitude modulation with a sinusoidal 
variation over the rotation cycle is consistent with a pure dipole pulsation mode, 
leading to the conclusion that the spots are not concentric with the magnetic 
field. While small departures between the spots and magnetic poles do 
occur in some magnetic Ap stars, e.g.\ in the roAp star 
HR\,3831 \citep{kurtzetal92},  the case of \acir\ requires equatorial spots far 
from the pulsation pole. For further progress, 
new, very precise measurements of the magnetic field of \acir\ 
are needed to show a convincing rotational variation, as we pointed out in 
Sect.~\ref{sec:rot}.

\subsection{Is \fonebold\ a dipole or a quadrupole mode?
\label{sec:dipole}}

The observed pulsation amplitude of an obliquely pulsating nonradial 
mode varies with rotation, as the aspect of the mode changes. This gives rise to a 
frequency multiplet, separated by the rotation frequency, that describes the 
amplitude modulation (and also phase modulation, if present). Within the formalism 
of the oblique pulsator model (\citealt{shibahashi86}; \citealt{kurtz86}; 
\citealt{kurtz90})  we have calculated the rotational amplitude modulation 
expected for three possible pulsation modes, ($l = 1, m = 0$), ($l = 1, m=\pm1$), 
and 
($l = 2, m= 0$), such that the amplitude modulation agrees with the observed 
amplitudes 
of the rotationally split frequency triplet given in Table~4. 
The results are shown in Fig.~\ref{fig:dipole} and each case will be discussed in 
detail below.

\subsubsection{\label{sec:l1m0}An axisymmetric dipole mode ($l=1,m=0$)}
 
\begin{figure} 
%\begin{center} 
\hskip -0.2cm
\includegraphics[width=8.8cm]{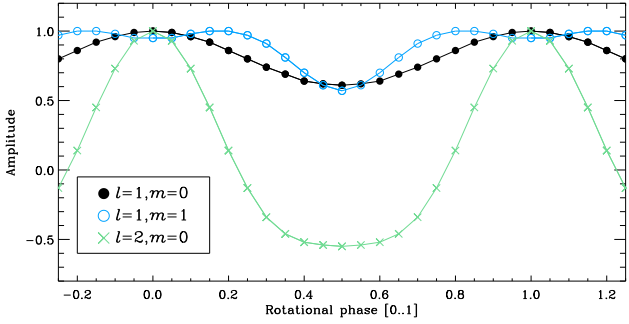} 
\caption{The relative amplitude modulation over the rotation cycle for 
two dipole modes, 
$l = 1, m = 0$ (solid circles) for $\beta = 18^\circ$ and 
$l = 1, m = 1$ (open circles)  for $\beta = 108^\circ$, and 
the axisymmetric quadrupole mode with 
$l = 2, m = 0$ (crosses) for $\beta = 51^\circ$ (in all cases $i=37^\circ$). 
The values of $\beta$ have been derived from the constraints of the 
amplitude ratio of the rotational sidelobes to the amplitude of the principal 
frequency
as described in detail in the text. 
For each mode the maximum amplitude has been normalised to 1.} 
\label{fig:dipole} 
%\end{center} 
\end{figure}

Whatever the relationship between the rotational light curve and the pulsation, 
the frequency triplet of Table~\ref{tab:sidelobes} is consistent with an 
axisymmetric dipole mode, or any mode of higher degree to which distortion by the 
magnetic field could result in the addition of a significant ($l=1$, $m=0$) 
component. This leads to the constraint (\citealt{kurtz90}): 
\begin{equation} 
\frac{A_{+1} + A_{-1}}{A_0} = \tan i \tan \beta = 0.246 \pm 0.014 = r_1, 
\end{equation}

\noindent where $i$ is the rotational inclination and $\beta$ is the magnetic 
obliquity of the negative pole, 
and $A_{+1}$, $A_{-1}$ and $A_0$ refer to the amplitudes of the 
rotational sidelobes and the principal frequency. Since we have already determined 
the rotational inclination to be $i = 37 \pm 4^\circ$ in Sect.~\ref{sec:rot}, that 
leads to $\beta = 18 \pm 3^\circ$. These values are consistent with 
only one magnetic pole being visible over the 
rotation cycle, i.e.\ with $i + \beta < 90^{\circ}$ as we have discussed in 
Sect.~\ref{sec:rot}. 
Fig.~\ref{fig:dipole} shows an axisymmetric dipole mode with these values
of $\beta$ and $i$ (solid circles). It has a sinusoidal amplitude modulation, 
which is in agreement with the observed modulation in Fig.~\ref{fig:ampvar}.

The oblique rotator model for a dipole magnetic field leads 
to a similar constraint on the geometry; \cite{hubrig07} found $\tan i \tan \beta 
= 0.66$ with no estimated error. We cannot, therefore, assess whether there is an 
inconsistency between these derivations of $\tan i \tan \beta$, but note that 
\cite{hubrig07} find $i = 38^\circ$, $\beta = 40^\circ$, hence $i + \beta < 
90^\circ$, also indicating that only one magnetic pole is visible. 

%%% _____________________________________________________________________
%%%
%%% Figure made with HB's program: dwkplot.pro on brixx' computer! 
%%% _____________________________________________________________________
%%%

\subsubsection{A non-axisymmetric dipole mode ($l=1,m=\pm1$)}
 
Non-axisymmetric dipole modes ($l = 1, m = \pm$1) 
within the oblique pulsator model constraints on the geometry (\citealt{kurtz86}, 
\citealt{shibahashi86}) give 

\begin{equation} 
\frac{A_{+1} + A_{-1}}{A_0} = -\frac{\tan i}{\tan \beta} = r_1, 
\label{eq:ratio} 
\end{equation} 

\noindent which leads to $\beta = 108\pm3^\circ$ for $i = 37\pm4^\circ$.  
This implies that both magnetic poles should be observed, contrary to 
observations. 
The amplitude modulation of a sectoral $l = 1, m=1$ mode with these
values of $\beta$ and $i$ is shown in 
Fig.~\ref{fig:dipole} (open circles) and exhibit a double wave that is not 
observed; 
this is because both poles are seen, but from different aspect. 
The pulsation mode can therefore not be a dipole mode with $m = \pm 1$. 
 
\subsubsection{An axisymmetric quadrupole mode $l = 2, m = 0$}
 
From our theoretical modelling in Sect.~\ref{sec:models} 
we show that both even and odd $l$ modes are present, 
so it is worth exploring whether the principal mode 
could have even degree, seen by the observer as a quadrupole mode. 
For an axisymmetric quadrupole mode with $l = 2, m = 0$ we expect a frequency 
quintuplet with separations of $f_{\rm rot} = 2.5841$\,$\mu$Hz. Since the outer 
two frequencies of this quintuplet are not observed, 
we assume that their amplitudes 
are below the 3$\sigma$ detection limit (i.e.\ $A_{\pm 2} < 0.018\pm0.002$\,mmag) 
and 
solve for the magnetic obliquity that is consistent with this. 
We then expect (\citealt{kurtz90}; \citealt{shibahashi86})

\begin{equation} 
\frac{A_{+1} + A_{-1}}{A_0} = \frac{12 \sin \beta \cos \beta \sin i \cos i}{(3 
\cos^2 \beta -1)(3 \cos^2 i -1)} = r_1, 
\label{eq:xx} 
\end{equation} 
and 
\begin{eqnarray} 
r_2 & \equiv & \frac{A_{+2} + A_{-2}}{A_0} \\ 
 & = & \frac{3 \sin^2 \beta \sin^2 i }{(3 \cos^2 \beta - 1)(3 \cos^2 i -1)} 
\label{eq:yy}\\ 
 & \le & 0.058\pm0.005. 
\end{eqnarray} 
We can divide Eq.~(\ref{eq:yy}) by Eq.~(\ref{eq:xx}) to yield 
\begin{equation} 
\tan i \tan \beta = \frac{4 r_2}{r_1} \le 0.94\pm0.10. 
\label{eq:beta} 
\end{equation} 
 
\noindent For $i = 37\pm4^\circ$, Eq.~(\ref{eq:beta}) yields $\beta = 51 \pm 
5^\circ$. 
This also fulfills the requirement that $i + \beta < 90^\circ$, but the amplitude 
variation is clearly not sinusoidal with rotation, as can be seen in 
Fig.~\ref{fig:dipole} (cross symbols). In fact, for this quadrupole the amplitude 
goes through 
zero and reverses phase over the rotation cycle, in clear conflict with the 
observations. We therefore conclude that the principal frequency does not have the 
observational properties of an axisymmetric quadrupole mode. 
 
\subsection{Concluding remarks on the mode identification of \fonebold}
 
The apparent lack of any rotational phase variation (Sect.~\ref{sec:prop})
and the rotational amplitude variation of the principal pulsation mode 
(Sect.~\ref{sec:dipole})
strongly support the supposition that the principal mode {\it has the 
observational properties of an axisymmetric $l = 1, 
m = 0$ dipole mode}. Note well, however, that we do not constrain our theoretical 
modelling in the next section to dipole modes for the principal pulsation mode, 
since magnetic perturbations may make modes of a different degree appear to be 
very 
similar to dipole modes observationally.

%---------------------------------------------------------------------------% 
% 
%			 	THEORETICAL MODELLING (Cunha, Brandao) 
% 
%---------------------------------------------------------------------------% 
\section{Asteroseismology of \acirbold} 

In the following we will compare theoretical pulsation models 
with the new observational constraints consisting of
the improved fundamental parameters (\paperi) and 
the detection of the large separation in the current work. 

\begin{figure} 
 \centering 
 \includegraphics[width=8.6cm]{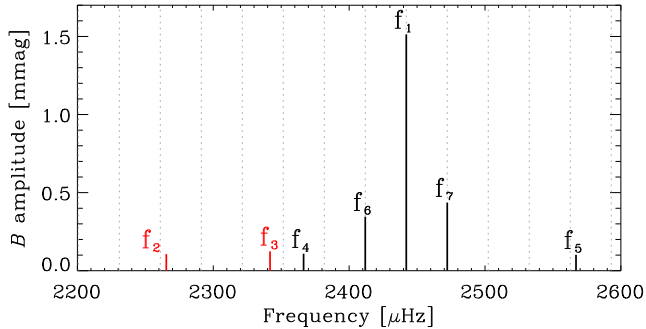} 
 \caption{Frequencies detected in \acir\ from the \wire\ data and 
$f_2/f_3$ from Kurtz et al.\ (1994). The vertical dashed lines mark the mean 
separation between $f_6$, $f_1$ and $f_7$.} 
 \label{fig:freq} 
%% HB: Program used to make this plot: acir/wire_acir_f_diagram.pro 
 \end{figure}

Before comparing the observed and theoretical frequencies 
it is crucial to include only the observed frequencies for which
there are no ambiguities about their detection. In Fig.~\ref{fig:freq} we 
show a schematic frequency spectrum of seven modes that we believe
are due to pulsation and intrinsic to \acir. 
In our list of secure frequencies in Table~\ref{tab:freq}
we have included the five frequencies $f_1$--$f_5$ detected by \cite{kurtz94}
and the two new frequencies $f_6$ and $f_7$. 
In Fig.~\ref{fig:freq} we have scaled the amplitudes to the $B$ filter system 
using the ratio $A_B / A_{\rm  WIRE}$ from Sect.~\ref{sec:saao}, 
and for $f_2$ and $f_3$ we use the scaling $A_v/A_B=1.29$ \citep{kurtz94}. 
The vertical dashed lines correspond to 
the average separation between consecutive 
modes in the triplet $f_6 + f_1 + f_7$.

The frequencies $f_2$ and $f_3$ were not detected in our data although
from the observed noise level in the \wire\ 2006 data and the observed
amplitude we should have detected them with a S/N ratio around 6--8.
A likely explanation is that their amplitudes have changed over time. 
Alternatively, it is possible that $f_2$ was not detected
due to its proximity to the sideband 
$f_2 \simeq f_1 - f_{\rm WIRE}$ ($\Delta f = 2.3$\,\mhz),
while the detection of $f_3$ was perhaps not possible
due to it being relatively close to an orbital 
harmonic ($n=13$, $\Delta f = 16.2$\,\mhz). 
These possibilities are illustrated in Fig.~\ref{fig:win}.

The frequencies $f_6$ and $f_7$ must have appeared some time between 
Feb'05 and Feb'06, since they were not detected in the \wire\ data sets from 
Sep'00 and Feb'05. If their amplitudes had not changed, the peaks should have been 
detected, with S/N of about $14$ and $7$ in the amplitude spectra from Sep'00 and 
Feb'05, respectively. Nevertheless, as they are seen in both of the simultaneous 
runs with \wire\ and from \saao\ in 2006, they must be intrinsic to \acir. 

A few additional frequencies have been reported based on spectroscopic campaigns 
\citep{balona03, kurtz06a}, but we have not included them in the modelling.
The frequencies reported by \cite{balona03} are clearly 
affected by 1 c\,d$^{-1}$  
aliasing problems, and the data analysis by \cite{kurtz06a} 
is complicated by the relatively poor frequency resolution 
while at the same time the frequencies lie close to the principal mode.

The frequencies shown in Fig.~\ref{fig:freq} contain information 
about the average properties of \acir\ as well as details about its internal 
structure and magnetic field. 
In the following we will attempt a forward comparison with the 
theoretical oscillation spectra obtained from models. To avoid confusion with the 
observed frequencies $f$, we shall designate the theoretical frequencies by 
$\nu_{n,l}$, where $n$ and $l$ are, respectively, the radial order and the degree 
of the eigenmodes.

% 
% ------------------------------------------------- 
\subsection{The large separation}
\label{sec:models}
% ------------------------------------------------- 
% 

The most commonly used observable quantity when studying 
high-overtone p~mode oscillations in stars is the large frequency separation 
$\Delta\nu_{n,l}$, defined as the difference 
between modes of the same degree and consecutive radial orders: 
$\Delta\nu_{n,l}\equiv \nu_{n+1,l}-\nu_{n,l}$. 
The large separation is a function of frequency, but for high 
radial orders the dependence is small.
In fact, in a spherically symmetric star, the frequencies of linear 
adiabatic acoustic pulsations of high radial order are expected to follow the 
asymptotic relation \citep{tassoul80}: 
\begin{equation} 
\nu_{n,l} \simeq \Delta\nu_0 \, (n + l/2 + 1/4 + \alpha) + \epsilon_{n,l}, 
\label{asymp} 
\end{equation} 
where $\Delta\nu_0=(2\int_0^R{\rm d}r/c)^{-1}$, $R$ is the stellar radius, $c$ is 
the sound speed and $r$ is the radial coordinate. The value of $\alpha$ depends on 
the properties of the reflection layer near the surface, and $\epsilon_{n,l}$ is a 
small term that depends mostly on the conditions in the stellar core. It can be 
seen that $\Delta\nu_0$ is the inverse of the sound travel time across the 
diameter of the star.

The high frequency of the oscillations observed in \acir\ indicates that these are 
indeed acoustic oscillations with high radial order. Thus one may expect to find 
in the oscillation spectrum a regular pattern corresponding to the asymptotic 
relation. 
However, the asymptotic analysis assumes spherical symmetry of the star. 
In \acir, as in other roAp stars, 
the presence of a strong magnetic field introduces a source of deviation from 
spherical symmetry that can shift the oscillation frequencies significantly 
from the asymptotic values \citep{cunha00, cunha06,saio04}. According to these 
studies, one may expect to still see regular frequency spacings in some parts
of the oscillation spectra, but abnormal spacing might also be found as in 
the well-studied case of HR~1217 (\citealt{kurtz05,cunha01}). 

The observed frequencies $f_6$ and $f_7$ lie almost equidistantly 
around $f_1$, to within a few nHz:
$f_1 - f_6 = 30.1746\pm0.0009\,\mu$Hz and $f_7 - f_1 = 30.1707\pm0.0005$\,\mhz. 
We determine the average separation from the slope
of a first order polynomial fitted to the three frequencies:
$\Delta f_{\rm obs} = 30.173\pm0.004\,\mu$Hz. 
To be conservative we adopt the difference between the two values
as the uncertainty. This separation is marked by the vertical
dashed lines in Fig.~\ref{fig:freq}. We note that the frequencies
$f_2$ and $f_5$ appear to roughly agree with the spacing.

If the regular frequency spacing of $f_6+f_1+f_7$ involves modes with degrees of 
different parity, 
and if their frequencies are not significantly modified by the magnetic field, 
then we may conclude that the large separation is $\Delta\nu_0\approx 2 \Delta 
f_{\rm obs}$. 
Otherwise, if only modes with degrees of the same parity are present, we have 
$\Delta\nu_0\approx \Delta f_{\rm obs}$. Since $\Delta\nu_0$ depends essentially 
on the mean density of the star, these two possibilities correspond to stars with 
very different mean densities. It is straightforward to determine, based 
on non-asteroseismic observables of \acir, which one of these possibilities 
is more plausible.

\begin{table} 
\caption{Global parameters of the CESAM model used to represent \acir. As input 
parameters we used $X_0= 0.70$, $Y_0=0.28$, $\alpha=1.6$ and no overshooting. 
$X_0$ and $Y_0$ are, respectively, the initial hydrogen and helium abundances and 
$\alpha$ is the mixing length parameter. The mass, $M$, luminosity, $L$, and 
radius, $R$, are expressed in solar units. $C_\nu$ is the homologous constant 
defined in Eq.~(\ref{eq:sep-scale-obs}).} 
\label{tab:models} \centering 
\begin{tabular}{ccccc} \hline \hline 
 $M$/M$_\odot$ & $L$/L$_\odot$ & $T_{\rm eff}$ [K] & $R$/R$_\odot$ & 
$C_\nu$ [$\mu$Hz] \\ \hline 
 1.715 & 1.019 & 7421 & 1.96 & 130.8 \\ 
 
\hline 
\end{tabular} \end{table}

\subsection{The pulsation model}
 
\begin{figure} 
 \centering 
 \includegraphics[width=8.2cm]{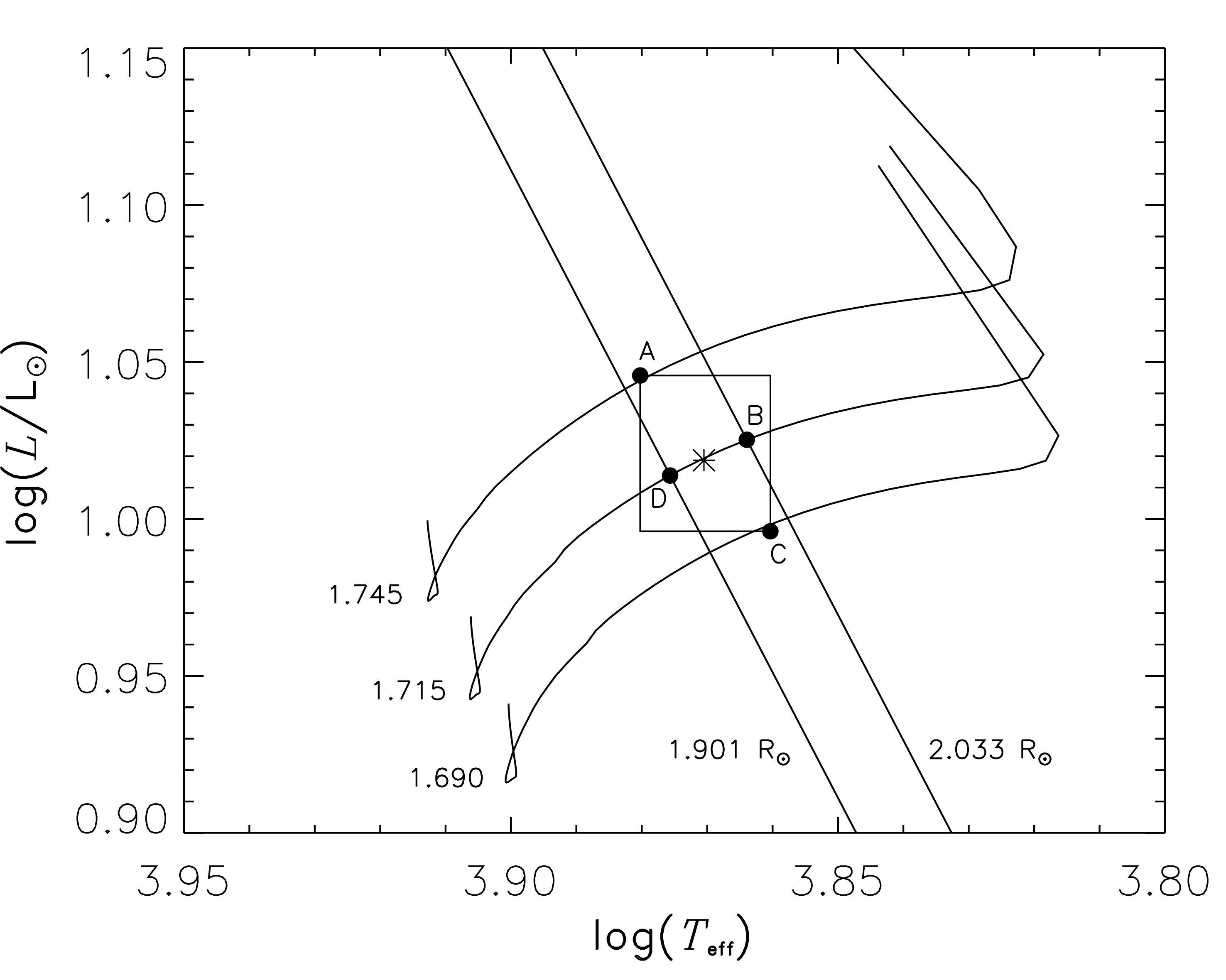} 
 \caption{H-R diagram with three evolution tracks from CESAM.
The position of our reference model is marked with an asterisk
and four additional models considered in our analysis 
are labeled A, B, C and D.
The rectangle marks the $1\,\sigma$ uncertainty on $T_{\rm eff}$ and
$L/{\rm L}_\odot$ and the two diagonal lines correspond to 
constant radii consistent with the $1\,\sigma$ uncertainty 
on the interferometric measurements from \paperi.} 
 \label{fig:hr} 
 \end{figure}

To compute theoretical pulsation frequencies we first calculated 
a reference equilibrium model of \acir\ using the stellar evolution 
code CESAM\footnote{The CESAM code is available at 

http://www.astro.up.pt/corot/models/cesam/} \citep{morel97}. 
The fundamental parameters used for this model are given in 
Table~\ref{tab:models} and are adopted from our analysis of 
interferometric data and spectra calibrated in flux as described in Paper~I. 
The position corresponding to these parameters is given by the asterisk 
in the Hertzsprung-Russell (HR) diagram in Fig.~\ref{fig:hr}. 
 
To calculate the theoretical oscillation frequencies for this model we used the 
linear adiabatic oscillation code ADIPLS \citep{jcd08}. 
Since the observed frequencies are above the acoustic cutoff frequency for 
the reference model, we have applied at the outermost point of the 
pulsation models a fully reflective boundary condition of the type $\delta p=0$, 
where $\delta p$ is the Lagrangian perturbation to the pressure. 
The physical mechanism responsible for the reflection of the oscillations in roAp 
stars has been under debate for decades. It is likely that the reflection of the 
oscillations results from the direct effect of the magnetic field \citep{sousa08}. 
Despite the small value of the longitudinal magnetic field of \acir\ 
(\citealt{hubrig04}; cf. Sec.~\ref{sec:rot}), 
we note that the magnetic field intensity of this star is not necessarily small. 
If we assume a dipolar topology and the magnetic obliquity and rotational 
inclination derived in  Sect.~\ref{sec:l1m0}, we find a polar magnetic strength of 
$\sim$ 1~kG. Moreover, if the magnetic field has a quadrupolar component, its 
polar intensity will be even stronger. 
A comparison between magnetic and gas pressures in the outer layers of a magnetic 
stellar model with
parameters similar to \acir\ was presented by \cite{cunhams07}. From their 
discussion
it is clear that even a 1~kG magnetic field has a strong effect on the dynamics 
of the oscillations throughout the whole atmosphere of a typical roAp star. 
Thus, it is expected that through mode conversion part of the mode energy is 
passed onto magnetic waves 
in the atmosphere. The energy transferred to these magnetic waves is kept in the 
mode. Thus, if the energy input in each cycle through the excitation mechanism 
\citep{balmforthetal01,cunha02,saio05} is enough to compensate for the fraction of 
the mode energy that is lost through running acoustic waves in the atmosphere 
\citep{sousa08} and 
through running magnetic waves in the interior \citep{robertsandsoward83}, the 
mode will be over-stable and may reach a detectable amplitude. 
 
The use of a fully reflective boundary condition is an artificial way to induce 
reflection of pulsations in the theoretical model and may shift the computed 
frequencies 
of individual modes away from their true values \citep{cunha06}. Nevertheless, its 
effect on the individual mode frequencies is expected to be systematic and, thus, 
should cancel out when computing pulsating quantities that are 
based on frequency differences, particularly when these differences involve only 
modes of the same degree. 
Since this is indeed the case for the large separation, 
our computed values will not be significantly affected 
by the adopted boundary condition.

%
% Figure prepared by MC + IB
% 
%-------------------------------------------------------------- 
% H-R diagram 
% 
%______________________________________________________________ 

The average large separation of the reference model 
is $\langle \Delta\nu_{n,l} \rangle=62.5$\,$\mu$Hz. 
Modes of degrees $l=0,1,2$ and 3, with frequencies around the observed values, 
were used to calculate this average. 
This value is close to $2\Delta f_{\rm obs}$ and so
the solution $\Delta\nu_0\approx \Delta f_{\rm obs}$ 
is firmly rejected.
This means that modes with degrees of alternating parity 
must be involved in the regular frequency pattern observed in 
the triplet $f_6$, $f_1$, $f_7$. 
 
It is beyond the scope of the present work to make a detailed model of \acir.
However, it is worth inspecting
how the results for our model compare with those for similar models placed within 
the region of uncertainty associated with the global parameters of \acir. We have
considered four additional models, labeled A, B, C and D in Fig.~\ref{fig:hr}. 
These models lie at the extremes of the $1\,\sigma$-uncertainty rectangle 
($T_{\rm eff}=7420\pm170$\,K, $L/{\rm L}_\odot=10.51\pm0.60$) 
and inside the strip defined by the two lines of constant radius 
corresponding to $R/{\rm R}_\odot = 1.967 \pm 0.066$ 
(the interferometric radius from \paperi). 
To estimate the large separations for 
these four models, we used the fact that the averaged asymptotic large separations 
follow a homologous relation of the type 
\begin{equation} 
\langle \Delta \nu_{n,l} \rangle = C_\nu \cdot 
 (M/{\rm M}_\odot)^{1/2}\; 
 (R/{\rm R}_\odot)^{-3/2}. 
 \label{eq:sep-scale-obs} 
\end{equation} 
The value of $C_\nu$ for our reference model of \acir\ is $130.8\,\mu$Hz.
In Table~\ref{tab:corners} we list the  masses of these models, as derived from 
\mbox{CESAM} evolution tracks using the same input parameters as for our reference 
model,
their radii, as derived from their luminosities and effective 
temperatures, and  the corresponding large separations, derived through the 
homologous scaling in Eq.~(\ref{eq:sep-scale-obs}).

\begin{table} 
\caption{Masses, radii and average large separations obtained from the homologous 
scaling 
given for models A, B, C, and D shown in Fig.~\ref{fig:hr}.} 
\label{tab:corners} \centering 
\begin{tabular}{lccc}     \hline \hline 
model & $M$/M$_\odot$ & $R$/R$_\odot$ & $\langle \Delta\nu_{n,l} \rangle$ 
[$\mu$Hz] \\ \hline 
A & 1.742 & 1.93 & 64 \\ 
B & 1.715 & 2.03 & 59 \\ 
C & 1.685 & 2.00 & 60 \\ 
D & 1.715 & 1.90 & 65 \\ 
 
\hline  
\end{tabular} \end{table} 

Inspection of Table~\ref{tab:corners} shows 
that the large separations of the models ($59$ to $65$\,\mhz) 
agree well with the observed value ($60$\,\mhz).
However, this is not the case for the high effective temperatures
that have been suggested in the literature (see \paperi\ for a discussion).
For an effective temperature of $7900$~K, and with a similar luminosity 
and mass, using the homologous scaling in Eq.~(\ref{eq:sep-scale-obs}) 
we find that the large separation of the model is about $75\,\mu$Hz, 
greatly exceeding the observed value. Hence, the asteroseismic modelling 
of the observed large separation strongly supports the effective temperature 
derived for \acir\ in \paperi.

\subsection{Evidence for magnetic perturbation of the modes} 

Despite the general agreement found between the observed 
and computed large separations, there are aspects 
about the observed frequencies that cannot be accounted for with the models 
considered here.
Firstly, only the frequencies $f_6$, $f_1$ and $f_7$ seem to follow a 
regular frequency pattern. In particular, the low amplitude mode $f_4$, which is 
seen in both 
the \wire\ data presented in this paper and the data acquired by \cite{kurtz94}, 
is offset by about a quarter of the large separation from where it would be 
expected, 
according to the asymptotic approximation.
Moreover, the almost-equal value 
of the separations ($f_1-f_6$) and ($f_7-f_1$) is also not reproduced by 
the reference model. If the triplet is assumed to correspond to modes 
of alternating parity and adjacent orders, 
then the difference between these two frequency separations can be written 
\begin{equation} 
\delta\nu_{n,l} \equiv 2\nu_{n,l}-\nu_{n-1,l+1}-\nu_{n,l+1}. 
\label{smallsep0} 
\end{equation} 
Under the same conditions as those required for Eq.~(\ref{asymp}) to be valid, 
$\delta\nu_{n,l}$ is asymptotically given by \cite[e.g.][]{cunha07}
\begin{equation} 
\delta\nu_{n,l}\simeq - 4( l + 1 ) {\Delta \nu_0 \over 4 \pi^2 \nu_{n,l}} 
\int_0^R {\dd c \over \dd r }{\dd r \over r } \; . 
\label{smallsep1} 
\end{equation} 
Given the increase of the sound speed with depth in the star, $\delta\nu_{n,l}$ is 
expected to be positive, increasing proportionally to $(l+1)$. 

% 
%-------------------------------------------------------------- 
% small separations (not the usual definition -- 
% see review Cunha et al 2007,eq.(37) 
 \begin{figure} 
 \centering 
 \includegraphics[width=8.6cm]{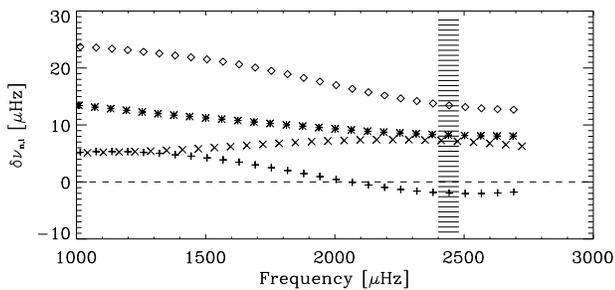}  
\caption{The frequency differences defined by Eq.~(\ref{smallsep0})
are shown versus mode frequency. Results are given for the cases 
when the triplet is assumed to involve modes of degrees 
$l=1,0,1$ (plus symbols), 
$l=2,1,2$ (crosses),
$l=3,2,3$ (asterisks), and 
$l=3,0,3$ (diamonds).
The observed value is marked by the horizontal dashed line and
the hatched region marks the location of the triplet.}
 \label{smallsep} 
 \end{figure} 
% -------------------------------------------------------------- 

The observed value for this quantity is 
$\delta\nu_{\rm obs} \equiv -((f_7-f_1)-(f_1-f_6)) = 0.0039 \pm 0.0014\,\mu$Hz.
In Fig.~\ref{smallsep} we plot $\delta\nu_{n,l}$ for our reference model 
(cf.\ Table~\ref{tab:models}), 
assuming different combinations of the
degrees of the modes in the triplet:
$l=(1,0,1)$, $l=(2,1,2)$, $l=(3,2,3)$, and $l=(3,0,3)$.
In the last case $\delta\nu_{n,l}$ is defined in a similar way, but 
with $(n-1,l+1)$ and $(n,l+1)$ replaced by $(n-2,l+3)$ and $(n-1,l+3)$, 
respectively. 
Except for the combination $l=(1,0,1)$, the values of $\delta\nu_{n,l}$ 
are positive and increase with $l$, as expected from the asymptotic expression. 
The deviation of $\delta\nu_{n,l}$ from the expected trend in the case 
of alternating modes of degrees $l=1$, $l=0$ and $l=1$ is not surprising, since in 
the 
asymptotic analysis leading to Eqs.~(\ref{asymp}) and (\ref{smallsep1}) it is 
assumed that the sound speed varies smoothly with depth. Because A-type stars have 
convective cores, that assumption breaks down at the edge of the core. The modes 
of lowest degree, which propagate deeper, will feel the effect of that rapid 
variation in the sound speed, and thus, their relative frequency is likely to 
deviate more strongly from the asymptotic behavior. 
 
Regardless of how closely the quantity $\delta\nu_{n,l}$ derived from our model 
follows the asymptotic behavior expressed by Eq.~(\ref{smallsep1}), it is clear 
that for all possible combinations of modes considered, this quantity is 
significantly different from the observed value 
(dashed horizontal line in Fig.~\ref{smallsep})
in the region where the modes are observed (the hatched region from $2.4$ to 
$2.5$\,mHz).
The smallest absolute value found for the theoretical 
$\delta\nu_{n,l}$ in the range of the observed frequencies, 
is around $2.0\,\mu$Hz for $l=(1,0,1)$, 
which is much larger than the $3.9$~nHz determined from the observations. 
Considering modes of degree higher than $l=3$ will not
solve this discrepancy, since $\delta\nu_{n,l}$ will increase with increasing $l$. 
Moreover, we find that this discrepancy is also present for models A, B, C, and D,
i.e.\ none of our theoretical calculations can explain the almost-equal 
separation observed between the three frequencies $f_6$, $f_1$ and~$f_7$. 
It remains to be investigated whether the effect of the magnetic field, 
which has been ignored in the models considered so far, may explain this 
and the other apparent disagreements between theoretical and observed quantities 
discussed here.

The effect of the magnetic field on the pulsations of roAp stars has been studied 
by different authors for over a decade 
\citep{dziembowski96,bigot00,cunha00,saio04,cunha06,sousa08}. 
However, it is only recently that magnetic models have started to be used, 
in a forward approach, in an attempt to fit the asteroseismic data of particular 
pulsators \citep{gruberbauer08,huber08}.  
The study of pulsations in the presence of a magnetic field is a complex one. 
In order to make it treatable, a number of assumptions and approximations have 
to be introduced. Thus, quantitative comparisons between observed frequencies 
and frequencies derived from magnetic models require some caution.  
Nevertheless, there are a number of robust results derived from the magnetic 
models that one may confidently use. \cite{cunha05} argued that the effect 
of a dipolar magnetic field is always to increase the large separation 
by a small amount, that can vary from nearly zero to a few $\mu$Hz. 
Since our non-magnetic models reproduce well the observed large separation, 
we expect that in the case of \acir\ the change in the large separation 
when including the magnetic field will be relatively small. 
On the other hand, a detailed magnetic modelling of \acir\ is necessary 
to test whether the effects produced by a magnetic field with properties 
consistent with the observational constraints may remove the discrepancies found 
between our models and the observed value of $\delta\nu_{n,l}$, 
as well as explain the deviation from the asymptotic 
trend of some of the observed oscillation frequencies.
 Such analysis will be presented in a separate paper (Brand\~ao et~al., in 
preparation).

Finally, we would like to emphasize that the almost-equal separation of the 
modes that compose the main triplet raises a question that goes beyond the simple 
issue of whether a particular magnetic field may bring the model predictions into 
agreement with the observations. In fact, even if such a magnetic field is found, 
its effect will depend on the frequency and so a precise cancellation of the 
difference $(f_7-f_1)-(f_1-f_6)$ may occur only in a very particular range of 
frequencies. Therefore, the question of whether there is an underlying reason for 
the principal modes to be found precisely in that range of frequencies, must also 
be considered. In particular, model simulations that provide a measurement of how 
often this precise cancellation is expected to occur need to be carried out, 
and compared with the empirical evidence from observations of other multi-periodic 
roAp stars.

\subsection{The \hipp\ and asteroseismic parallaxes} 
 
Prior to the \wire\ observations presented in this paper, the pattern of observed 
frequencies observed in \acir\ by \cite{kurtz94} suggested a large separation of 
50\,\mhz. 
\cite{matt99} used this value, together with an effective 
temperature of $T_{\rm eff}=8000$\,K and a mass of $M=2.0\pm0.5\,{\rm M}_\odot$ to 
calculate the expected ``asteroseismic'' parallax of \acir. 
The value they determined, $\pi_{\rm ast}=51.3\pm1.2$, was significantly 
smaller (7\,$\sigma$) than the parallax measurement by \hipp\ ($\pi_{\rm Hip-
97}=60.97\pm0.58$\,mas; \citealt{hipparcos97}). 
Moreover, a similar calculation for another 11 roAp stars indicated that the 
\hipp\, parallaxes were 
systematically larger than the parallaxes derived from the asteroseismic data of 
these 
roAp 
stars. The new, higher value for the large separation found in this paper, 
together with the lower mass and \teff\ from \paperi, resolve the discrepancy 
found for \acir. Using Eq.~(4) from \cite{matt99} we derive a luminosity of \acir\ 
of $L = 10.6 \pm 1.5$\,L$_\odot$. We adopt a mass of $M = 1.7 \pm 0.2$\,M$_\odot$ 
and $T_{\rm eff} = 7420 \pm 170$\,K (both values from \paperi), a bolometric 
correction of ${\rm BC} = 0.19\pm0.08$ (\citealt{brandao07}; average of their two 
estimates), ${\rm M}_{{\rm bol,}\odot}=4.75$, $V=3.19$, and $E(B-V)_{\rm 
max}=0.06\pm0.02$ (from \citealt{matt99}) to get $\pi_{\rm ast}=62.8\pm5.3$\,mas. 
This is 
in good agreement with the parallax from \hipp, which was recently updated to
$\pi_{\rm Hip-07}=60.36\pm0.14$\,mas\ \citep{leeuwen07}. This resolves the problem 
of the low asteroseismic parallax for \acir\ found by \cite{matt99}. Moreover, the 
lower effective temperature of \acir\ found in \paperi\ corroborates one of the 
suggestions of \cite{matt99}, namely that the systematic discrepancy between the 
asteroseismic and \hipp\ parallaxes could be associated with a systematic error in 
the determination of the effective temperatures of roAp stars. 
 
\begin{figure} 
 \centering 
 \includegraphics[width=8.8cm]{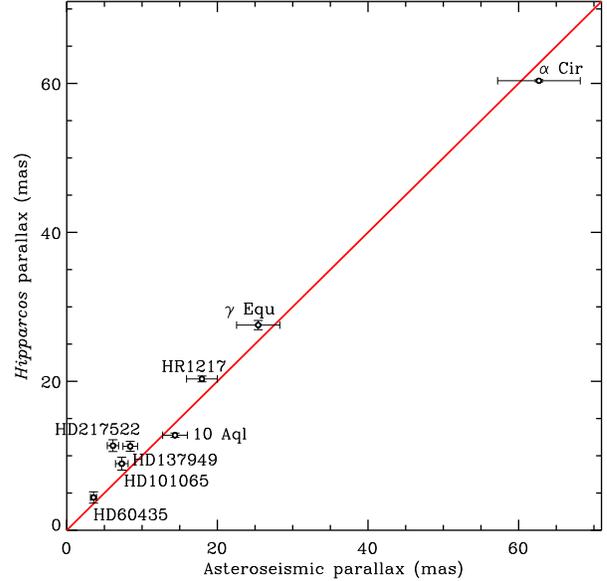} 
 \caption{The \hipp\ parallaxes of eight roAp stars are 
plotted vs.\ the asteroseismic parallaxes as determined from the large separation, 
\teff, and mass. 
This is an updated version of Fig.~1 from Matthews et al.\ (1999).
 \label{fig:matt}} 
 \end{figure}

In Fig.~\ref{fig:matt} we plot the \hipp\ parallaxes versus the asteroseismic
parallaxes for eight roAp stars. This is an update of the original Fig.~1 from 
\cite{matt99}. Four stars are omitted (HD 119027, 166473, 190290, and 203932) 
because the uncertainty on their \hipp\ parallax is 8--10 times larger than for 
the other stars. For \acir\ we used the values found above, and for the other 
stars we used the information in Tables~1 and 2 of \cite{matt99} and updated 
parallaxes from \cite{leeuwen07}. Note that we have adopted larger uncertainties 
on 
\teff\ ($\sigma_T=200$\,K) and on mass ($\sigma_M=0.5\,{\rm M}_\odot$). 
The solid line in Fig.~\ref{fig:matt} shows equality, and it seems that there is 
now acceptable agreement for most stars, the exceptions being HD\,217522 and 
HD\,137949. 

 % 
%-------------------------------------------------------------- 
% Program: .r wire_matt99_plot.pro 
% 
%______________________________________________________________ 

 %______________________________________________________________ 

% Program: .r wire_acir_f_diagram 
 
% 
%______________________________________________________________ 

%---------------------------------------------------------------------------% 
% 
%			 	CONCLUSION 
% 
%---------------------------------------------------------------------------% 
\section{Conclusions}\label{sec:concl} 

We have analysed observations of the roAp star \acir, 
comprising 84 days of photometric time series from 
the star tracker on the \wire\ satellite. 
We used data from four different runs 
lasting between 8 and 42 days collected over six years,
or nearly the entire mission life time of \wire\ \citep{bruntt07gott}.

The four light curves from \wire\ all show a double wave 
modulation with a period of $P_{\rm rot} = 4.4792\pm0.0004$\,d and 
a peak-to-peak amplitude of 4\,mmag. 
We interpret this variation as being due to spots 
on the surface giving us directly the rotation period of the star.
This is the first direct detection and was only made 
possible with the very stable photometry 
-- on both short and long timescales -- from \wire. 
Interestingly, the amplitude
and phase of the modulation did not change significantly 
from 2000 to 2006 and hence the spot configuration seems 
to be extremely stable.
The rotation period confirms an earlier indirect 
detection by \cite{kurtz94}. Our initial assessment is that the spots 
need to be equatorial to explain the double wave rotational light curve, hence -- 
unusually for an Ap star -- the spots are not associated with the single, visible 
magnetic pole. Further study of the rotational light variation is in progress.

At much shorter periods (7 minutes) we detect 
the known dominant mode of pulsation at 2442\,mHz;  
in addition, we have discovered two new frequencies
located symmetrically around the principal mode.
This triplet of frequencies is detected in the two runs from 2006, 
but not in the first two runs in 2000 and 2005.
This is not due to differences in detection sensitivity,
therefore the new frequencies 
must have appeared sometime between 2005 and 2006.
This new discovery is confirmed by our ground-based
photometry during 16 nights from \saao,
collected simultaneously with the two \wire\ runs in 2006.
The separation of the peaks is $\Delta f_{\rm obs} = 30.173 \pm 0.004$\,\mhz.
Based on our theoretical models of \acir\ we interpret 
this as half the large separation, meaning that the
triplet consists of modes with spherical degree of alternating parity.
To confirm that our interpretation is correct will require that 
one or more frequencies with this separation are observed. 

%% This may be possible with a high-speed spectroscopic campaign from the ground.

The principal mode has symmetrical sidelobes with a frequency
separation equal to the rotational frequency, $f_{\rm rot}=2.58$\,\mhz.
We find that at the rotational phase of pulsation amplitude maximum, 
the pulsation phases of the three frequencies 
are equal within uncertainties. 
In the framework of the oblique rotator model this
is expected when the pulsation axis and the rotation
axis are not aligned.
Thus, the three frequencies describe the change 
in observed pulsation amplitude as the observer 
views the deformation of the star from different perspectives.
We have used the ratio of the amplitudes of the rotational frequency triplet
to test different possibilities for the spherical degree
and azimuthal order of the principal mode. 
We conclude that only modes with observational 
properties similar to those of a mode with $l=1,m=0$ agree with the observations.
This could be the dipole mode itself, or it could be a mode of higher 
degree that is distorted by the magnetic field such that a dipolar component 
dominates the observed rotational amplitude modulation.

We have computed models that are consistent with the fundamental
parameters of \acir\, using as constraints the interferometric
radius and the effective temperature from \paperi. 
For the first time in the study of \acir, 
the theoretical large separations derived from the models are
in good agreement with the observations. 
The separations between the observed modes 
in the triplet $f_6+f_1+f_7$ differ by only 3.9~nHz and
this cannot be reconciled with the theoretical calculations presented in this 
work.
We have discussed that magnetic effects may possibly account for 
this discrepancy but postpone the detailed analysis to a future 
publication (Brand\~ao et~al., in preparation).

We have compared the \hipp\ parallax of \acir\ 
with the value inferred from the large separation,
assuming our interpretation above is correct. 
We find good agreement and this seems to resolve
the issue of the discrepant value for \acir\ as discussed by \cite{matt99}.

%---------------------------------------------------------------------------% 
% 
%                           ACKNOWLEDGEMENTS 
% 
%---------------------------------------------------------------------------% 
 
\section*{Acknowledgments} 
 
This research has been funded by the Australian Research Council. 
DWK acknowledges support for this work from 
the UK Science and Technology Facilities Council (STFC). MC is supported by the 
Ciencia2007 Programme from FCT (C2007-CAUP-FCT/136/2006) and by FCT 
and FEDER (POCI2010) through the projects PDCTE/CTE-AST/81711/2006 and 
PTDC/CTE-AST/66181/2006. IMB is supported by FCT through the grant 
SFRH/BD/41213/2007. 
GH is supported by the Austrian "Fonds zur F\"orderung der wissenschaftlichen 
Forschung" under grants P18339-N08 and P20526-N16.
IMB is very grateful to Jo\~ao Pedro Marques for providing useful material 
in relation to the theoretical models. We thank Vladimir Elkin for
providing a high-precision measurement of \vsini\ for \acir.
Part of the this work was 
supported by the European Helio- and Asteroseismology Network (HELAS), 
a major international collaboration funded by the European Commission's 
Sixth Framework Programme.
This research has made use of the SIMBAD database, operated at CDS, Strasbourg, 
France. 
 
%---------------------------------------------------------------------------% 
% 
%                               REFERENCES 
% 
%---------------------------------------------------------------------------% 
  
\bibliography{bruntt_wire_acir} 

\bsp 
\label{lastpage} 
 
\end{document}